\documentclass{jetpl}
\usepackage{graphicx}
\usepackage{amsmath}
\usepackage{amssymb}
\usepackage{hyperref}
\newcommand{\DD}{\,\mathrm{d}}
\newcommand{\st}[1]{\ensuremath{\mathit{I}[#1]}}
\newcommand{\pd}[2]{\ensuremath{\frac{\partial #1}{\partial #2}}}
\twocolumn

\title{Spin diffusion in liquid $^3$He confined in nafen}

\rtitle{Spin diffusion in liquid $^3$He confined in nafen}

\sodtitle{Spin diffusion in liquid $^3$He confined in nafen}

\author{V.\,V.\,Dmitriev\thanks{e-mail: dmitriev@kapitza.ras.ru},\,L.\,A.\,Melnikovsky,\,A.\,A.\,Senin,\,A.\,A.\,Soldatov,\,A.\,N.\,Yudin}

\rauthor{V.\,V.\,Dmitriev,\,L.\,A.\,Melnikovsky,\,A.\,A.\,Senin,\,A.\,A.\,Soldatov,\,A.\,N.\,Yudin}

\sodauthor{V.\,V.\,Dmitriev,\,L.\,A.\,Melnikovsky,\,A.\,A.\,Senin,\,A.\,A.\,Soldatov,\,A.\,N.\,Yudin}

\address{P.\,L. Kapitza Institute for Physical Problems RAS,
2 Kosygina str., 119334 Moscow, Russia}

\dates{15 May 2015}{*}

\begin{document}


\abstract{
We report results of spin diffusion measurements in normal phase of liquid
$^3$He confined in nafen. Nafen is a new type of aerogel and it consists of
Al$_2$O$_3$ strands which are nearly parallel to one another at macroscopic distances.
We examine two samples of nafen with different porosities using spin echo techniques.
Spin diffusion of $^3$He along and across the strands was measured.
The aerogel alignment is clearly evident from observed spin diffusion anisotropy.
A theory describing this effect is developed and compared with the experiment.
}

\maketitle

\section{Introduction}
Superfluid $^3$He in high porosity aerogel is a model system to
investigate the influence of impurities on unconventional
superfluidity.
Silica aerogels which consist of a nearly chaotic array of SiO$_2$ strands
are used
in most of such experiments. An important
parameter for theoretical models of superfluid $^3$He in aerogel is the
mean free path ($\lambda$) of Fermi-liquid quasiparticles which can be
determined from measurements of spin diffusion coefficient ($D$). At high
temperatures ($T\gtrsim 20$\,mK) the density of quasiparticles is large, so
$\lambda$ and $D$ follow the bulk Fermi-liquid behavior, i.e.
$\lambda\propto T^{-2}$ and $D\propto T^{-2}$. At low enough temperatures
($T<10$\,mK) aerogel strands limit the mean free path and the spin
diffusion coefficient, so that at $T\sim 1$\,mK the density of quasiparticles is so small that values
of $\lambda$ and $D$ are fully determined by the array of aerogel strands
and do not depend on $T$. Such behavior was observed in the first
measurements of spin diffusion in $^3$He confined in nearly isotropic
95\% and 98\% open silica aerogels \cite{Cand,Bunkov}. In some recent
experiments with superfluid $^3$He in aerogel a new type of aerogel was
used \cite{we12,we14,Bph}. The remarkable feature of this aerogel, called
``nematically ordered'' (N-aerogel), is that its strands are oriented
along the same direction. There are two types of N-aerogel: ``Obninsk
aerogel''
produced by Leypunsky Institute (Obninsk, Russia) \cite{oxidmelt}
which consists of AlOOH strands
and nafen \cite{nafen} which consists of Al$_2$O$_3$ strands.
In the limit of $T=0$ the strong global
anisotropy of N-aerogel should result in anisotropy of $^3$He spin
diffusion. For example, in ``Obninsk aerogel'' with overall density
$\sim30$\,mg/cm$^3$ the spin diffusion along the strands
is about twice as fast as that in perpendicular direction \cite{previous}.

Here we present results of the spin diffusion measurements in $^3$He
confined in nafen, which is much denser than ``Obninsk aerogel'' and
compare the obtained results with the theory we developed.

\section{Theory}
An approximate description for the weak field spin diffusion in bulk $^3$He
was first given by D.\,Hone~\cite{hone}
\begin{equation}\label{Dhone}
j_\mathbf{M}^l =- \frac{v_F \lambda}{3}(1+F^{a}_0) \pd{\mathbf{M}}{x^l},
\end{equation}
where $\mathbf{M}$ is the magnetization, $\mathbf{j}_\mathbf{M}$ is respective current,
$v_F$ is the Fermi velocity, $F^{a}_0$ is the Landau Fermi-liquid parameter,
and $\lambda$ is a quasiparticle mean free path.
Diffusion in anisotropic aerogel is described by a generalized equation
\begin{equation}\label{anisotropic_fick}
j_\mathbf{M}^l =- D^{lm}\pd{\mathbf{M}}{x^m},
\end{equation}
where $D^{lm}$ is the spin diffusion tensor.
It is important to note, that no mean free path can characterize diffusion
in anisotropic medium.
Diffusive flux \eqref{anisotropic_fick} in such system depends
on the gradient direction and they are not necessarily parallel to each
other. On the other side, a mean free path is a scalar, it is by its
nature averaged over quasiparticle distribution which is isotropic
even in anisotropic aerogel.

Suppose that the stationary kinetic equation
\begin{equation}\label{boltzmann}
\pd{n}{x^m}\, \pd{\epsilon}{p^m} -
\pd{n}{p^m}\, \pd{\epsilon}{x^m} =
\st{n},
\end{equation}
holds separately for each individual spin component.
In the left hand side of the kinetic equation \eqref{boltzmann} a local
equilibrium function should be substituted. This gives~\cite{hone}
\begin{equation*}
\pd{n}{x^m}\, \pd{\epsilon}{p^m} -
\pd{n}{p^m}\, \pd{\epsilon}{x^m} =
-\psi^m \pd{n_0}{\epsilon}\pd{\epsilon}{p^m},
\end{equation*}
where
\begin{equation*}
\psi^m=\left(1+F_0^a\right)\frac{2\pi^2\hbar^3}{p_F m^*}\pd{M}{x^m}.
\end{equation*}
The magnetization here and below is taken in magneton units $\gamma\hbar=1$.

Calculation of $D^{lm}$ is greatly simplified at low temperature
so that the collisions between Fermi quasiparticles can be neglected. This means that
the scattering of quasiparticles on the aerogel alone is responsible
for the collision integral
\begin{equation*}
\st{n}=
\int\DD\mathbf{p}'
\left(
w(\mathbf{p},\mathbf{p}') n' (1-n)-
w(\mathbf{p}',\mathbf{p}) n (1-n')
\right),
\end{equation*}
where $n=n(\mathbf{p})$, $n'=n(\mathbf{p'})$, and
$w(\mathbf{p},\mathbf{p}')=w(\mathbf{p}',\mathbf{p})$ is the
scattering $\mathbf{p'} \rightarrow \mathbf{p}$ probability.
Since the scattering by aerogel strands conserves the spin and the energy, the collision
integral vanishes for equilibrium distribution function of the true
quasiparticle energy $n_0(\epsilon)$.
It is therefore possible to linearize the collision integral
using $\delta\tilde{n}=n-n_0(\epsilon)$
\begin{equation*}
\st{n}=
\int\DD\mathbf{p}'
w(\mathbf{p},\mathbf{p}')
\left(\delta\tilde{n}'-
\delta\tilde{n}
\right).
\end{equation*}
Furthermore, the solution of the kinetic equation can be sought for in the form
\begin{equation*}
\delta\tilde{n}=\pd{n_0}{\epsilon}\chi(\hat{\mathbf{p}}),
\quad
\hat{\mathbf{p}}=\mathbf{p}/p,
\end{equation*}
where $\chi$ depends on the direction of $\mathbf{p}$ only.
The kinetic equation becomes
\begin{equation}\label{kinetic_use}
-\psi^m \pd{\epsilon}{p^m}=
v_F
\int
\DD\sigma (\hat{\mathbf{p}},\hat{\mathbf{p}}')
(\chi'-\chi),
\end{equation}
where $\DD\sigma$ is the differential scattering cross section per unit volume.

The flux $\mathbf{j}_\mathbf{M}$ also vanishes for $n_0(\epsilon)$
and can be linearized as follows
\begin{equation}\label{flux}
j_\mathrm{M}^l=
\int\frac{\DD\mathbf{p}}{(2\pi\hbar)^3}
\pd{\epsilon}{p^l}\delta\tilde{n}=
-\frac{p_F^2}{(2\pi\hbar)^3}
\int\DD\mathbf{\hat{p}}
\chi(\hat{\mathbf{p}})
\hat{p}^l.
\end{equation}

We now have all tools ready for the diffusion tensor
calculation. The differential cross section $\DD\sigma$ depends
on the microscopic aerogel structure and the properties of quasiparticle
scattering by aerogel strands. Let us represent the aerogel as
an array of infinite cylindrical strands with diameter $d$ oriented in
$z$ direction and randomly distributed in $xy$ plane
with the surface density $N$. The porosity of such structure
is therefore
\begin{equation*}
p=1-\frac{\pi N d^2}{4}.
\end{equation*}
For the scattering on the cylinder walls we
consider two opposite limits: diffuse and specular reflection.

\subsection{Diffuse Reflection}
After diffuse reflection all information about the velocity direction of the incident
particle is lost. The scattering cross section on a unit wall area is
\begin{equation*}
\DD\sigma_\mathbf{m}=
-\frac{\DD \hat{\mathbf{p}}'}{\pi}
\begin{cases}
0,								&(\mathbf{m} \hat{\mathbf{p}})>0;\\
0,								&(\mathbf{m} \hat{\mathbf{p}}')<0;\\
(\mathbf{m} \hat{\mathbf{p}}) (\mathbf{m} \hat{\mathbf{p}}'),	&\text{otherwise,}
\end{cases}
\end{equation*}
where the unit vector $\mathbf{m}$ is the outer normal to the surface element.
For a system of $z$-aligned cylinders, this vector is uniformly distributed in $xy$ plane.
The differential scattering cross section for this system is obtained by integration
\begin{equation}\label{sigma_diffuse}
\begin{split}
\DD\sigma&=
\frac{Nd}{2}\int\DD\mathbf{m}\, \delta(m_z) \DD\sigma_\mathbf{m}= \\
&=\frac{Nd}{4\pi}\sin\theta \, \sin\theta' \,
\bigl|
\sin \Delta -
\Delta \cos \Delta
\bigr|\DD\hat{\mathbf{p}}'
,
\end{split}
\end{equation}
where $\Delta=\phi-\phi' \in (-\pi, \pi)$, and the spherical angles
$(\theta,\phi)$ and $(\theta',\phi')$ correspond
to $\hat{\mathbf{p}}$ and $\hat{\mathbf{p}}'$ respectively.

Generally, the diffusion tensor $D^{lm}$ has three principal values.
Due to the axial symmetry of the system,
it can be characterized by the mere two distinct components
$D^\parallel$ and $D^\perp$. It is therefore sufficient to calculate the diffusion in
two directions: along and across the aerogel axis.
In the former case the kinetic equation \eqref{kinetic_use} is
\begin{equation}\label{kin13}
-\psi\cos\theta
=\int (\chi'-\chi)\DD\sigma.
\end{equation}
If we substitute \eqref{sigma_diffuse} in \eqref{kin13} and use the fact
that the distribution function does not depend on
the polar angle $\phi$, we get
\begin{equation*}
-\frac{\pi\psi\cos\theta}{2Nd}
=\int (\chi'-\chi)
\sin\theta \, \sin^2\theta' \,
\DD\theta'
.
\end{equation*}
The solution of this equation
\begin{equation*}
\chi=
\frac{\psi}{Nd}
\frac{\cos\theta}{\sin\theta}
\end{equation*}
should be substituted in \eqref{flux} to get the flux
\begin{equation*}
\begin{split}
j_\mathrm{M}&=
-\frac{p_F^2 \psi}{(2\pi\hbar)^3 Nd}
\int\DD\mathbf{\hat{p}}
\frac{\cos^2\theta}{\sin\theta}= \\
&=-\frac{p_F^2 \psi}{8\pi\hbar^3 Nd}=
-D_\text{D}^\parallel \pd{M}{z}
,
\end{split}
\end{equation*}
where
\begin{equation}\label{DDparall}
D_\text{D}^\parallel=
\left(1+F_0^a\right)
\frac{\pi v_F}{4 Nd}=
\frac{\pi^2 \left(1+F_0^a\right)
v_F}{16}
\frac{d}{1-p}
.
\end{equation}

Similar procedure can be used to investigate the lateral diffusion.
If $\psi^m$ has only $x$ component, then
the kinetic equation has the form
\begin{equation*}
-\psi\sin\theta\,\cos\phi
=\int (\chi'-\chi)\DD\sigma.
\end{equation*}
It has a solution
\begin{equation*}
\chi=\frac{16 \psi}{Nd \left(\pi^2+16\right)}\cos\phi,
\end{equation*}
which leads to the magnetization flux
\begin{equation*}
\begin{split}
j_\mathrm{M}^l&=-\frac{p_F^2}{(2\pi\hbar)^3}
\frac{16 \psi}{Nd \left(\pi^2+16\right)}
\int\DD\mathbf{\hat{p}}
\cos^2\phi
\sin\theta= \\
&=-D_\text{D}^\perp
\pd{M}{x}
,
\end{split}
\end{equation*}
where
\begin{equation}\label{DDperp}
\begin{split}
D_\text{D}^\perp&=
\left(1+F_0^a\right)\frac{2\pi v_F}{\left(\pi^2+16\right) Nd}= \\
&=\frac{\pi^2 \left(1+F_0^a\right) v_F }{2 \left(\pi^2+16\right)}\frac{d}{1-p}.
\end{split}
\end{equation}
The diffusion anisotropy is described by the ratio
\begin{equation}\label{DD}
D_D^{\parallel}/D_D^\perp=(\pi^2+16)/8\approx3.23\,.
\end{equation}

\subsection{Specular Reflection}
Elementary scattering cross section on a unit smooth surface is
\begin{equation*}
\DD\sigma_\mathbf{m}=
-\DD \hat{\mathbf{p}}'(\hat{\mathbf{p}}\mathbf{m})
\begin{cases}
0,								&(\mathbf{m} \hat{\mathbf{p}})>0;\\
\delta\left(
    \hat{\mathbf{p}}'-\hat{\mathbf{p}}+2\mathbf{m}(\hat{\mathbf{p}}\mathbf{m})
    \right)								&\text{otherwise.}
\end{cases}
\end{equation*}
For the array of cylinders we again integrate over possible $\mathbf{m}$ orientations
\begin{equation}\label{sigma_specular}
\begin{split}
\DD\sigma&=
\frac{Nd}{2}\int\DD\mathbf{m}\, \delta(m_z) \DD\sigma_\mathbf{m}= \\
&=\frac{Nd}{4}
\delta(\theta-\theta')
\,
\left|
\sin \frac{\phi - \phi'}{2}
\right|
\DD\hat{\mathbf{p}}'
.
\end{split}
\end{equation}
The delta function guarantees the conservation
of the $z$ component of momentum. This means that strictly
parallel specular cylinders do not obstruct the spin
flow along the system axis. Corresponding component
of the diffusion tensor $D_\text{S}^\parallel$ (infinite in our model)
is limited either by strand irregularities or quasiparticle-quasiparticle collisions.
Here we formally calculate only lateral (finite) component
of the diffusion tensor.

We again write down the kinetic equation
\begin{equation*}
-\psi\sin\theta\,\cos\phi
=\int (\chi'-\chi)\DD\sigma.
\end{equation*}
It has a solution
\begin{equation*}
\chi=
\frac{3\psi}{4Nd}
\cos\phi,
\end{equation*}
which results in the magnetization flux
\begin{equation*}
j_\mathrm{M}^l=
-
\frac{3\psi}{4Nd}
\frac{p_F^2}{(2\pi\hbar)^3}
\int
\DD\phi
\DD\theta
\sin^2\theta
\cos^2\phi
=
-D_\text{S}^\perp
\pd{M}{x}
,
\end{equation*}
where
\begin{equation}\label{DSperp}
D_\text{S}^\perp=
\left(1+F_0^a\right)
\frac{3\pi v_F}{32Nd}=
\frac{3\pi^2
\left(1+F_0^a\right)
v_F }{128}
\frac{d}{1-p}.
\end{equation}

\section{Details of experiments}
We have used two samples of nafen with overall densities 90 and
243\,mg/cm$^3$ (nafen-90 and nafen-243). Their structure was
investigated in \cite{we15}. The samples consist of strands with diameter
of $\approx8$\,nm and $\approx9$\,nm correspondingly. The density of bulk
Al$_2$O$_3$ is 4\,g/cm$^3$, this gives for the porosities of the samples 97.8\%
(nafen-90) and 93.9\% (nafen-243).

The experimental chamber used in the present work was similar to the
chamber described in \cite{we12}. It is made of Stycast-1266 epoxy
resin and has two separate cells. The cuboid samples
with characteristic sizes of 4\,mm are placed freely in the cells, so
that $\sim70$\% of each cell is filled with nafen.

Experiments were carried out using spin-echo technique in magnetic field $\approx280$\,Oe (corresponding NMR frequency is $\approx900$\,kHz) at pressure of 2.9\,bar.  In order to
avoid a paramagnetic signal from solid $^3$He on the surface of strands, the samples were
preplated by $\sim 2.5$ atomic layers of $^4$He. The strands of the samples were oriented parallel to the external steady magnetic field $\bf H$. Two gradient coils
were used to apply the field gradient in directions parallel and perpendicular to the strands.
The necessary temperatures were obtained by nuclear
demagnetization cryostat and were measured by a quartz tuning fork.
The temperature was determined in assumption that the resonance
linewidth of the fork in normal $^3$He is inversely proportional to
the temperature \cite{fork}. We calibrated the fork at high temperature
where the diffusion coefficient in aerogel should be
the same as in bulk $^3$He (using \cite{bulk} as a reference).

Spin echo decay curves were obtained by standard two-pulse method:
we measured amplitude of the echo after $\pi/2 - \tau - \pi/2$
pulses, where $\tau$ is the delay between pulses. The measurements
were carried out at temperatures from 1.4\,mK up to 60\,mK for two directions of magnetic field gradient
(parallel and perpendicular to the direction of aerogel strands)
and at several values of the gradients ($0.24\div 1.25$\,Oe/cm).

\section{Results}
Expression for the spin echo amplitude
\begin{equation}\label{ampl}
I=I_{0}\exp(-2\tau/T_{2}-A\tau^{3})
\end{equation}
can be found from Bloch-Torrey equations~\cite{Abragam}. With an
obvious generalization for an anisotropic media, the coefficient
$A$ here is given by
\begin{equation}
A=\frac{2}{3}\gamma^2 D^{lm} G^l G^m,
\end{equation}
where $\gamma$ is the gyromagnetic ratio, $G^l$ is the magnetic field
gradient.

\begin{figure}[t]
\center
\includegraphics[width=0.95\linewidth]{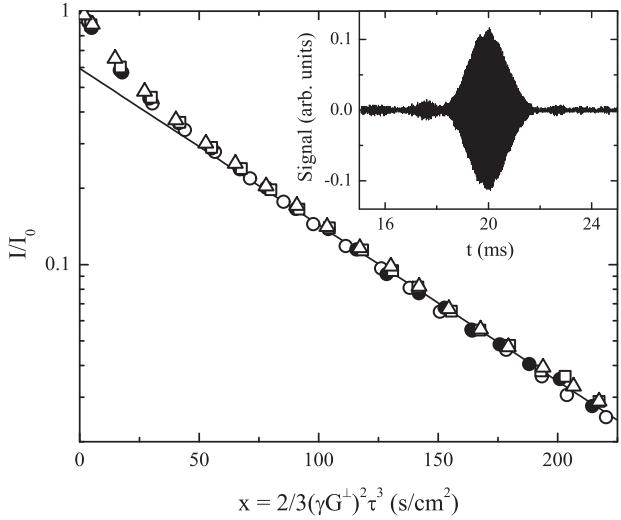}
\caption{Fig. \thefigure:
Spin echo decay in $^3$He in nafen-90 for different field gradients applied in direction perpendicular to aerogel strands. $G^\perp=0.70$\,Oe/cm (open circles), 0.53\,Oe/cm (filled circles), 0.37\,Oe/cm (open squares), 0.24\,Oe/cm (open triangles). $H=278$\,Oe. $T\approx 2.0$\,mK. Solid line is the best fit of the data at $x>100$ by Eq.(\ref{ampl}).
Inset: Typical echo signal of $^3$He in nafen-90. $T\approx 2.0$\,mK, $G^\perp=0.53$\,Oe/cm, $\tau=10$\,ms.}
\label{echo}
\end{figure}

Typical echo signal of $^3$He in nafen is shown in the inset in
Fig.\ref{echo}. In order to determine the value of spin diffusion
coefficient, spin echo amplitudes should be measured for different $\tau$
and then fitted by Eq.(\ref{ampl}). In this procedure the term with $T_2$
can be neglected,
because observed relationship between $I/I_0$ and $G^2\tau^3$ does not depend on field
gradient at all used temperatures. An example data set is shown in Fig.\ref{echo}.
Note that at $x\leq80$ experimental points in Fig.\ref{echo} deviate from the
linear dependence. It is due to the presence of bulk $^3$He outside the aerogel sample. At low temperatures the spin diffusion in bulk $^3$He is greater
than that in aerogel. However, for the same reason the relative contribution
of bulk $^3$He into the total echo signal rapidly decreases with the
increase of $\tau$, so we determined the value of spin diffusion
coefficient $D(T)$ of $^3$He in aerogel from the data at relatively
large $x$ where they follow the linear dependence ($x>100$ for
Fig.\ref{echo}).

\begin{figure}[t]
\center
\includegraphics[width=0.93\linewidth]{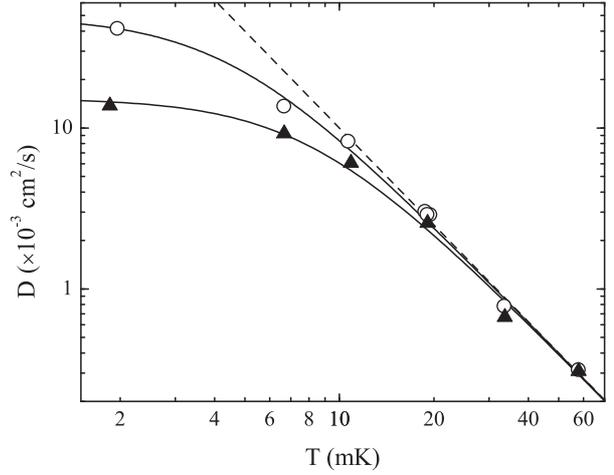}
\caption{Fig. \thefigure:
Temperature dependence of the spin diffusion tensor:
$D_{90}^\parallel(T)$ (open circles)
and $D_{90}^\perp(T)$ (filled triangles) in nafen-90.}
\label{diff1}
\end{figure}
\begin{figure}[t]
\center
\includegraphics[width=0.93\linewidth]{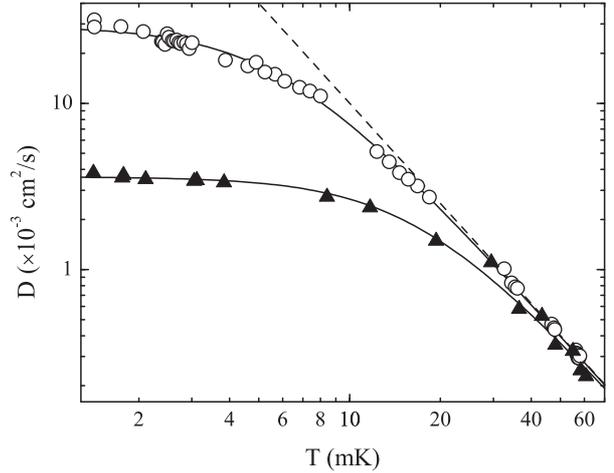}
\caption{Fig. \thefigure:
Temperature dependence of the spin diffusion tensor:
$D_{243}^\parallel(T)$ (open circles)
and $D_{243}^\perp(T)$ (filled triangles) in nafen-243.}
\label{diff2}
\end{figure}

The measured temperature dependencies $D(T)$ for two
orientations of the gradient are shown in Fig.\ref{diff1} for nafen-90
and in Fig.\ref{diff2} for nafen-243. In order to obtain a value of spin diffusion coefficient
in zero temperature limit ($D\equiv D(0)$) these dependencies were fitted by the equation
\begin{equation}\label{difsum}
    D^{-1}(T)=D_{b}^{-1}(T)+D^{-1},
\end{equation}
where $D_b\propto T^{-2}$ is the diffusion coefficient in bulk $^3$He, which is determined only by collisions between quasiparticles.
Solid lines in these graphs are fits by Eq.(\ref{difsum}),
dashed lines -- diffusion coefficient in bulk $^3$He (extrapolation to $P=2.9$\,bar using the data presented in \cite{bulk}).

Obtained principal values of the spin diffusion tensor are:
\begin{itemize}
\item in nafen-90
\begin{align*}
&D_{90}^\parallel \approx 0.049\,\text{cm}^2/\text{s},\\
&D_{90}^\perp     \approx 0.015\,\text{cm}^2/\text{s},\\
&D_{90}^\parallel / D_{90}^\perp \approx 3.3.
\end{align*}
\item in nafen-243
\begin{align*}
&D_{243}^\parallel \approx 0.029\,\text{cm}^2/\text{s},\\
&D_{243}^\perp     \approx 0.0036\,\text{cm}^2/\text{s},\\
&D_{243}^\parallel / D_{243}^\perp \approx 8.1.
\end{align*}
\end{itemize}
We estimate the accuracy of these values as $\pm 10$\%.

\section{Discussion}
In isotropic system the spin diffusion coefficient
is (see Eq.(\ref{Dhone}))
\begin{equation}\label{D0}
    D=\frac{v_F \lambda}{3}v_F(1+F^{a}_0).
\end{equation}
As explained in Section 2 no mean free path can properly
characterize the case of globally anisotropic aerogel. However, to
account for the experimental data it is convenient to introduce zero-temperature effective
mean free paths $\lambda_{\parallel}$ and $\lambda_{\perp}$ using
Eq.(\ref{D0}) as the definition.
If  $v_F=5397\,\text{cm}/\text{s}$ and $F_0=-0.717$
\cite{3he} then for nafen-90 we get
$\lambda_{\parallel}\approx 960\,\text{nm}$,
$\lambda_{\perp}\approx 290\,\text{nm}$
and for nafen-243
$\lambda_{\parallel}\approx 570\,\text{nm}$,
$\lambda_{\perp}\approx 70\,\text{nm}$.

In our experiments the strands of nafen were preplated with $^4$He. In
this case the reflection of quasiparticles on the surface is believed
to be specular or at least partly specular \cite{fre, FR}. For specular
reflection $D_S^{\parallel}$ is limited by
imperfect strands alignment and irregularities on their surface.
Therefore we can not directly compare the measured ratio $k=D^{\parallel}/D^\perp$
with proposed theory for specular case.
However, for nafen-243 we obtain $k\approx8.1$ which is much
greater than it is expected for diffuse one (see Eq.(\ref{DD})).
This means that the reflection in nafen-243 is predominantly specular.

The values of $D_D^\perp$ and $D_S^\perp$ can be calculated from
Eqs.(\ref{DDperp}), (\ref{DSperp}) using values of $\rho$ and $d$ for our samples. These theoretical values should weakly depend on
variations in orientations of the strands and can be compared with
experiment. For the case of nafen-90 we obtain
$D_S^{\perp}\approx 0.013\,\text{cm}^2/\text{s}$ and
$D_D^{\perp}\approx 0.010\,\text{cm}^2/\text{s}$ while the experimental value
$D_{90}^\perp\approx 0.015\,\text{cm}^2/\text{s}$.
For nafen-243
$D_S^{\perp}\approx 0.0052\,\text{cm}^2/\text{s}$,
$D_D^{\perp}\approx 0.0043\,\text{cm}^2/\text{s}$
and the experimental value
$D_{243}^\perp\approx 0.0036\,\text{cm}^2/\text{s}$.

We can conclude that the experimental values qualitatively agree with the
theoretical model. The quantitative difference may be due to experimental
errors in determination of $D$, errors in nafen parameters, and
variations of strands diameters (by $\pm 20$\% as follows from
\cite{we15}).

\section{Acknowledgements}
We are grateful to I.M.\,Grodnensky for providing the samples of nafen and to E.V.\,Surovtsev for useful discussions.
This work was supported in parts by RFBR (grants 13-02-00674, 13-02-00912), Russian Science Support Foundation
and the Basic Research Program of the Presidium of Russian Academy of Sciences.

\end{document}